\documentclass[10pt, conference, compsocconf]{IEEEtran}

\usepackage{amsthm}
\usepackage[cmex10]{amsmath}
\usepackage{amsfonts}
\usepackage{amssymb}

\ifCLASSINFOpdf
  \usepackage[pdftex]{graphicx}
  \usepackage{subfigure}
\else

\usepackage{graphicx}
\usepackage{subfigure}
\fi
\usepackage{epstopdf}

\usepackage{algorithmic,algorithm}

\usepackage[numbers,sort&compress]{natbib}
\usepackage{color}

\usepackage{url}

\usepackage{mdwtab}

\begin{document}

\title{Co\_Hijacking Monitor: Collaborative Detecting and Locating Mechanism for HTTP Spectral Hijacking}


\author{\IEEEauthorblockN{Pan Wang}
\IEEEauthorblockA{College of Internet of Things\\Nanjing University of Posts\&Telecommunications\\
Emails: wangpan@njupt.edu.cn }
\and
\IEEEauthorblockN{Xuejiao Chen}
\IEEEauthorblockA{Department of Communication \\Nanjing college of Information Technology \\
	Emails: chenxj@njcit.cn }}

\maketitle

\begin{abstract}
With the rapid growth of mobile internet, mobile application, like website navigation, searching, e-Shopping and app download, etc. are all popular in worldwide. Meanwhile, it become more and more popular that traditional HTTP protocol, which is also applying in not only web browsing but also communication between mobile application clients and servers. Besides, it has made HTTP Hijacking profitable. Furthermore, it has brought a lot of troubles for users, network operators and ISP. We analyze the principle of HTTP spectral Hijacking and present a mechanism of collaboratively detecting and locating called Co\_HijackingMonitor. Experimental result shows that, Co\_HijackingMonitor can solve the hijacking problem effectively. 
\end{abstract}


\section{Introduction}\label{sec:intro}
In recent years, website navigation, search, e-commerce, mobile application download are widely adopted and HTTP protocol is becoming more and more popular. HTTP protocol is not only used in web browsing, but also a large number of communications between mobile application clients and servers use HTTP. 

This has brought huge underground market space using HTTP hijacking technology. Besides, it also brought great trouble to  users, network operators and internet service providers. Hijacker can hide on the way of all routing nodes. The means of hijacking are endless and continue to retrofit, such as homepage modifying, process hooking, system boot hijacking, LSP injection, browser plug-in hijacking~\cite{browserhijacking}, HTTP proxy filter~\cite{httpproxy}, kernel data packet hijacking~\cite{kernelhijacking}, bootkit~\cite{bootkit}, CDN cache pollution~\cite{cachepollution}. 

Because network operators own the network link, they have inherent advantages for HTTP hijacking. The usual way is spectral hijacking bypassing communication link. From the point of view of end-to-end, HTTP hijacking could occur in the side of internet terminal, network link and cloud. This article specifically aimed at the research of  HTTP hijacking and detecting technology on the network link( or called spectral hijacking). HTTP hijacking, in short, is an attack to monitor HTTP message in a dedicated data channel established by the user and the web server. Once meeting the default conditions, the hijacker will insert the well-designed network data packets into the normal data stream. The purpose is to allow the client program to interpret the ‘error' data and pop up a new window in the web browser to display the contents of promotional ads or directly display a web site in order to make a large profit. 

Several typical methods of HTTP hijacking~\cite{kong2014}:
(1)Hijackers add or tamper the channel code of Ad network or website promotion. They mainly aimed at website navigation, web search engine and some major e-commerce website. The hijackers use HTTP hijacking technology to modify or add promotional channel code brutally in order to grab the profit of target website promotion. Such hijacking applications have been extended to web navigation and searching, software and mobile application (APK/EXE) downloading, game distribution, and so on. 
(2)Hijackers add extra advertisements on normal sites, like pop-up windows.

HTTP hijacking has caused varying degrees of harm to both users and service providers. For users, it is not only easy to leak the user's sensitive information, but also hijack a large number of internet user's behavior. Finally,it is very easy to cause the risk of online financial transactions. For service providers, a large number of promotional costs were grabed. At the same time, HTTP hijacking can also make user's web browsering fail or redirect.  Therefore, the detection and prevention of HTTP hijacking are of great significance.

The first section of this paper introduces the related work.  In the second section, the principle of HTTP hijacking on the network link is analyzed in detail. The third section describes the design scheme of the Co\_Hijacking Monitor mechanism in detail; including TTL detection based on dynamic updating method of target website routing hash table, HTTP response behavior, cooperative monitoring and traceability. In the fourth section, the experiment is carried out in a real network. We evaluate the ability of Co\_Hijacking to detect and locate the HTTP spectral hijacking attack. 

\section{related work}
In recent years, both the industry and academic pay much attention to the problem of HTTP hijacking on the operator's network links.However, it is still very difficult for network operators to detect HTTP spectral hijacking in time. The existing detection schemes are either not practical enough or lack of efficiency. Details are below.

In this paper, we think that we should start from the principle of hijacking if we plan to find a practical and effective method of detecting HTTP spectral Hijacking. We should find out the rule of the HTTP spectral hijacking and then build the model of spectral hijacking originally. In addition, there are a large number of links in the communication network.That is to say, hijacker can attack from a lot of nodes of network, so it is necessary to use cooperative method of monitoring in order to cover all the possible attack points. From the point of view of the target website, the related work in the field of spectral hijacking can be classified into two categories:
One is DNS hijacking, the other is HTTP hijacking. The DNS hijacking occurs frequently on the network link, because DNS server is more concentrated, attackers is easy to deploy attacking devices. So this kind of attacks have frequently emerged since 2003, until 2015 when Baidu sued China Chongqing Telecom DNS hijacking, which made it reach the peak~\cite{chongqing}.  
The detection and location of DNS hijacking can be divided into three types:
Passive monitoring detection, false packet detection and cross check query. 
The method of passive monitoring is mainly to capture all DNS requests and responses pairs, and then detect hijacking attacks by behaviors analysis~\cite{yanboru2016}. False message detection using the initiative to send a probe to detect whether there is a DNS hijacking~\cite{yanboru2016,kong2014}. Cross check query is mainly to further reverse the query of the results of DNS analysis, and to detect hijacking attacks by determining whether the two result of query are consistent. 
Each of these methods has its advantages and disadvantages, in which the passive monitoring detection is a passive way, the other two are active. Passive monitoring detection will not cause the network's additional traffic, but the method cannot apply to the new attack or variant means with the change of DNS hijacking behavior. False packet detection requires a large number of active packets to send, it will increase extra burden over the network. Cross detection query greatly depends on the capabilities of DNS server with reverse query service , but a lot of DNS does not have the ability. Because HTTP spectral hijacking is a kind of new attack in recent years, the related research work is less accumulated. Network operators mostly use the source address detection method to detect whether there are packets with fake addresses in the network~\cite{wangpeng2013,yangbo2015}. 
But because there are a lot of other network attacks, such as DDoS, which also uses a fake address, so this method cannot be applied to identify and detect the HTTP hijacking accurately. 

Some researchers use flow identification methods to detect such attacks~\cite{Hajikarami2014,Meda2014,Mauro2014}. However, this kind of method is often unable to achieve accurate identification. So it is a lack of operability in real network.

\section{HTTP hijacking principle analysis}
\subsection{Spectral hijacking deployment structure}
\begin{figure}[ht]
	\centering
	\begin{tabular}{cc}
		\includegraphics[width=2.8 in]{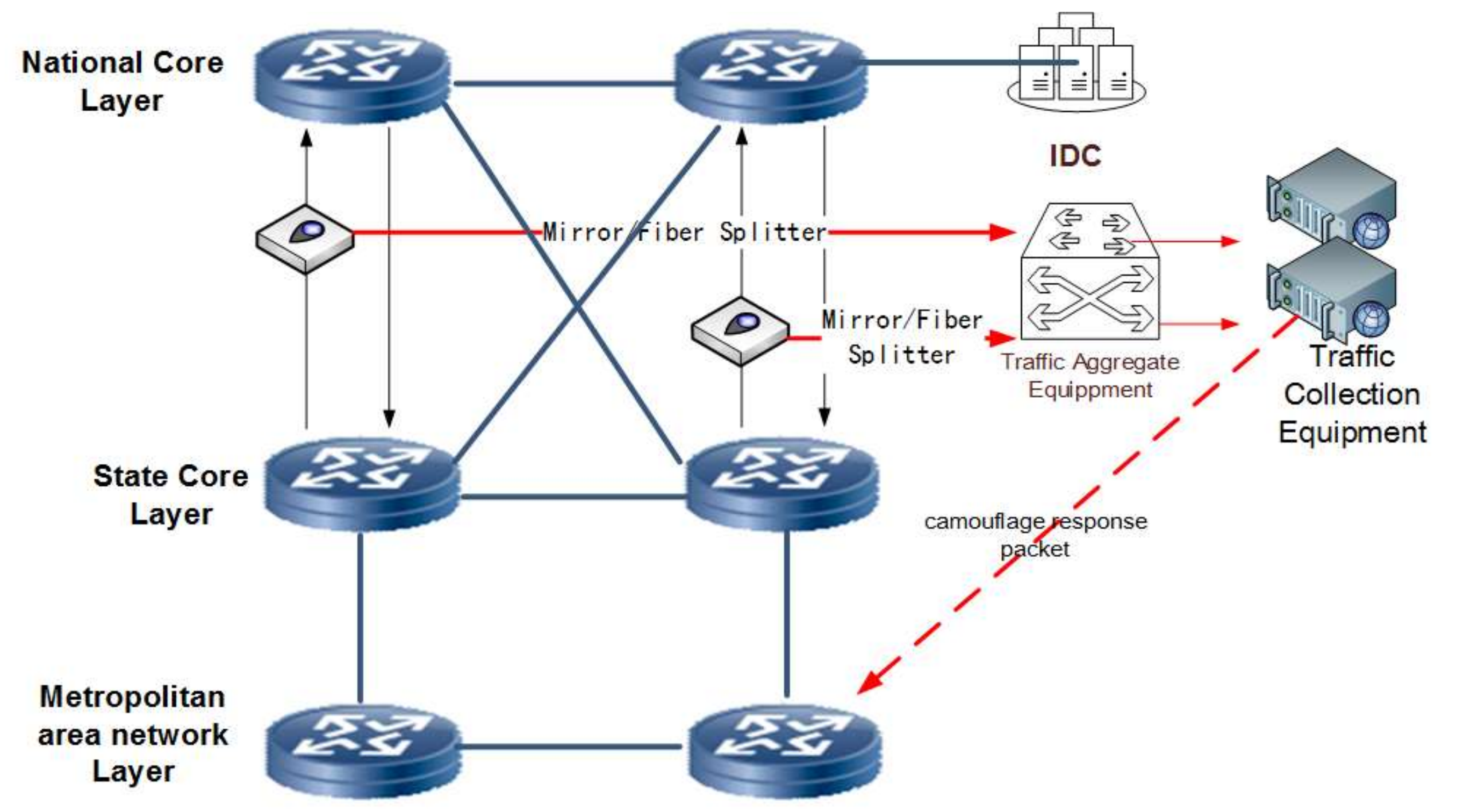}
	\end{tabular}
	\caption{spectral hijacking topology on network link}\label{fig:Figure1hijackingdeployment}
\end{figure}

As shown in Fig.~\ref{fig:Figure1hijackingdeployment}, under usual circumstances, the attacker sniffs all the spectral traffic information by deploying the related traffic acquisition device at the core router. There are other circumstances where operators deploy DPI/CDN/ cache and other systems in bypass which hijack attackers use to launch attacks. The acquisition device can be configured by policy, only collect part of traffic such as DNS, HTTP uplink GET requests, etc.  When a normal HTTP session request is routed through the device, the acquisition device uses the HTTP hijacking technique to hijack the HTTP session, and implement the next attack, such as sending camouflage response packet to the hijacked client in order to force the client to jump to a false site, etc. 

\subsection{HTTP hijacking principle}
After the web server that is accessed by a user's browser and sends an HTTP request, routers of the operator will first receive the HTTP request, the bypass device of the operator routers also received a copy of the HTTP request after spectral or mirroring.  Then send camouflage response packet for hijacking before the web server returning the data.  The browser will jump to the target web site after receiving the camouflage response, the real data of the web server will be discarded after the true data of the web server. 
\begin{figure}[ht]
	\centering
	\begin{tabular}{cc}
		\includegraphics[width=2.8 in]{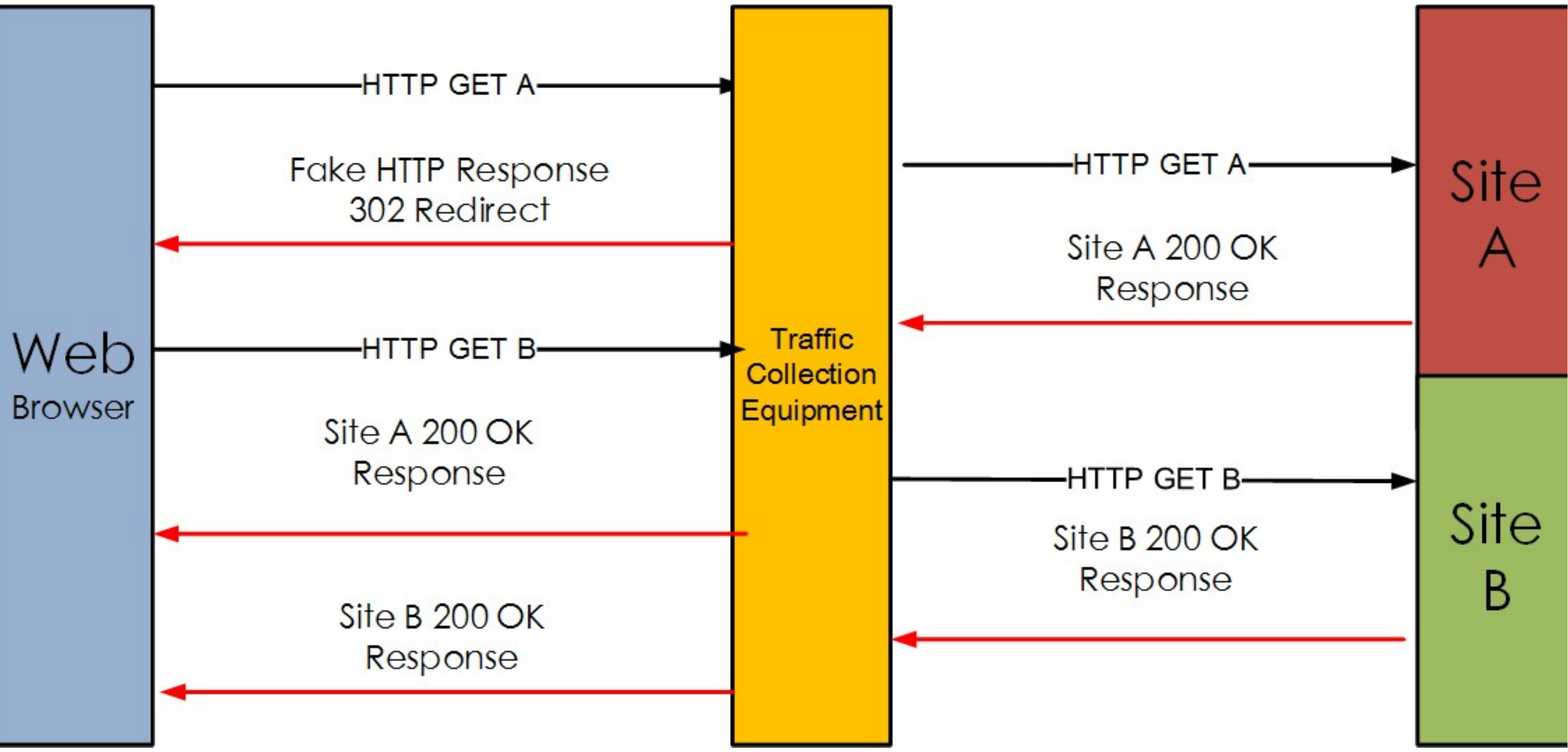}
	\end{tabular}
	\caption{HTTP response 302 redirect hijacking schematic}\label{fig:Figure2redirect302}
\end{figure}
\begin{figure}[ht]
	\centering
	\begin{tabular}{cc}
		\includegraphics[width=2.8 in]{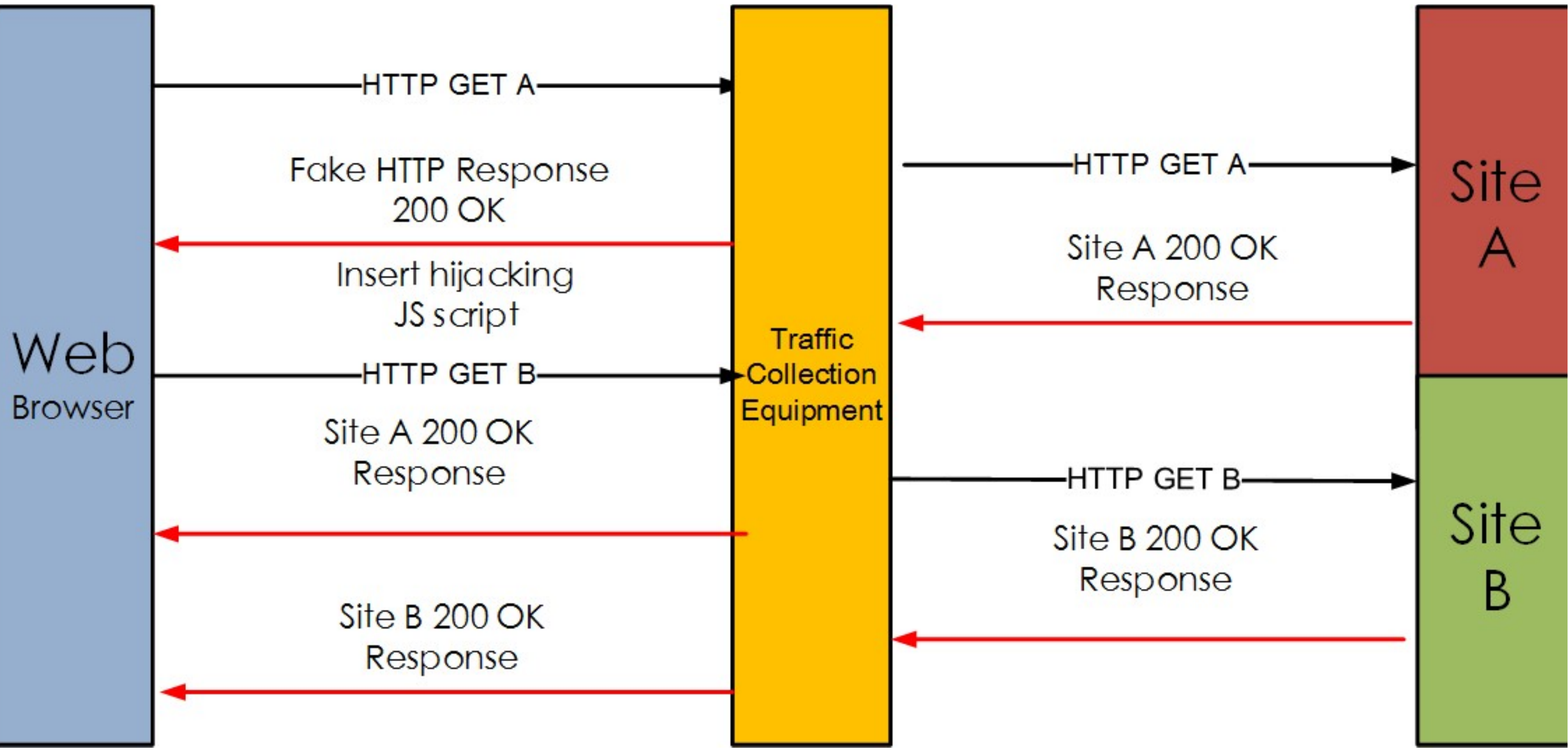}
	\end{tabular}
	\caption{HTTP response 200 OK hijacking schematic}\label{fig:Figure3OK200}
\end{figure}

According to the different disguised response package, can be divided into 302 redirect hijacking and 200 OK hijacking method. As shown in Fig.~\ref{fig:Figure2redirect302} and Fig.~\ref{fig:Figure3OK200}. HTTP 302 redirect hijacking, as its name implies, that is, in the way of HTTP 302 redirect response to disguise the return of the response packet, and fill in the hijacking target site URL in the jumping URL in the internal packet, in order to achieve the purpose of forcing users to visit the target site B browser.  HTTP 200 OK hijacking is to use HTTP 200 OK response to camouflage response packet, and inject a script in the JS script, in order to achieve forced users to display the browser hijacking side display content, such as pop, floating window, Banner , other advertising content. 

\subsection{The law of HTTP hijacking}
\begin{figure}[ht]
	\centering
	\begin{tabular}{cc}
		\includegraphics[width=2.8 in]{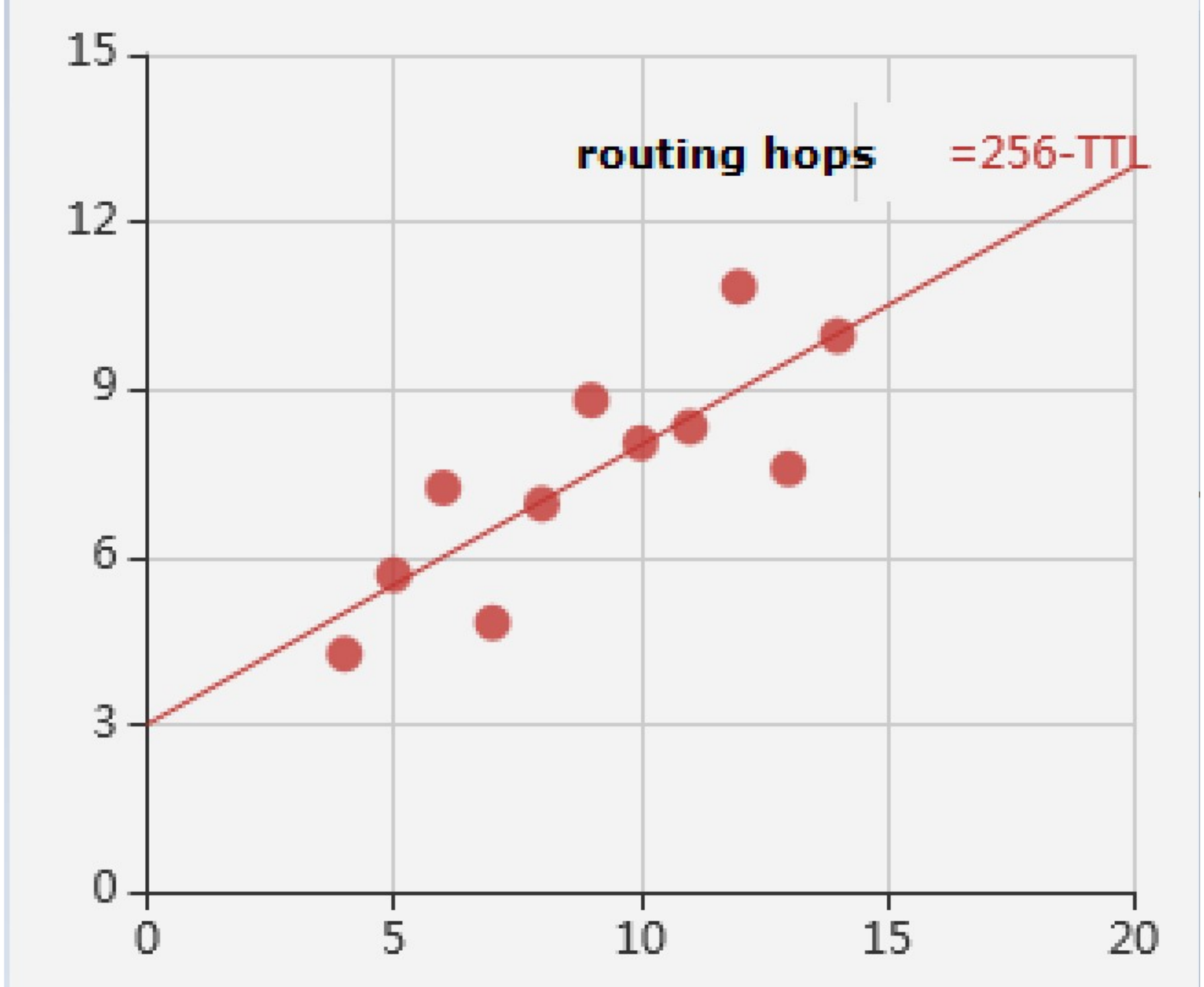}
	\end{tabular}
	\caption{The trend of the number of route hops in normal session}\label{fig:rhops1}
\end{figure}
\begin{figure}[ht]
	\centering
	\begin{tabular}{cc}
		\includegraphics[width=2.8 in]{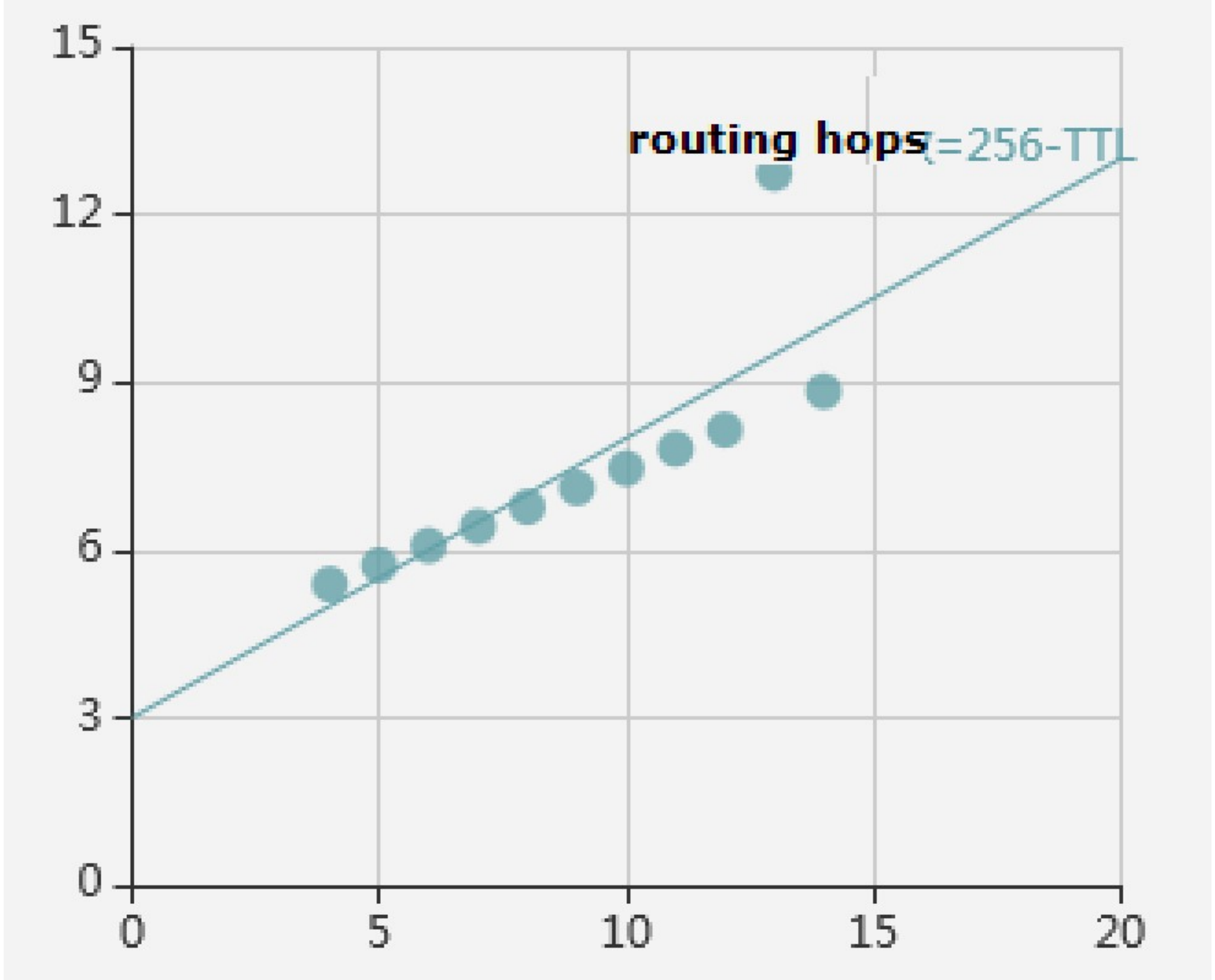}
	\end{tabular}
	\caption{The trend of the number of routes in the hijacking session}\label{fig:rhops2}
\end{figure}

From the above principle, we can see that the HTTP hijacking has the following characteristics:
\subsubsection{TTL value} The TTL value of the camouflage response packet is often less than the normal value, mainly due to the hijacking device in the network link is located between  the target site and users, so the TTL value is less than normal, and the TTL value is the same packet camouflage, because spectral hijack points are fixed, the camouflage response to the user's routing number is basically the same. As shown in Fig.~\ref{fig:rhops1} and Fig.~\ref{fig:rhops2}.
\subsubsection{HTTP 302 redirect response} The proportion of HTTP 302 redirect packets in the network is higher than the normal value. 
\subsubsection{sequence number} When the hijacking occurs, the hijacked session often has a request that corresponds to one of the previous two different response packets, and the sequence number of the two response packets is the same.

\section{MECHANISM DESIGN}
\subsection{Detection mechanism description}
We designed a new method for real time detection of spectral hijacking-Co\_Hijacking Monitor based on the principle of the HTTP and the description of the law in the second quarter.  First, the mechanism is modeled from the TTL value, or the number of routing hops, and the subsequent behavior of the normal response of the HTTP, study the rule of HTTP, mine out the difference between the HTTP spectral hijacking and the normal conversation behavior, and detect the TTL value for each HTTP session through the establishment of a set of large-scale dynamic updates of the target site routing hash table.

Meanwhile, for the part of the attackers, may use the TTL value to avoid tampering detection, use HTTP subsequent normal response behavior to determine whether it is a hijacking behavior.  Third, select operator network access layer BRAS level to carry out traffic spectral detection in order to cover as much as possible the detection range of the attack.  Then set up a whole set of the whole network spectral monitoring system, detect HTTP spectral hijacking behaviors efficiently and quickly, it also puts forward the means of tracing the point of attack.

The Co\_Hijacking Monitor mechanism proposed in this paper has the following 4 characteristics:
\subsubsection{impact on the network} Access layer on the operator's network, in addition to collecting the HTTP uplink and downlink handshake information, does not need to make any changes to the existing network infrastructure, that facilitates the practical application and incremental deployment. 
\subsubsection{TTL tampering} Judge comprehensively whether the existence of hijacking by TTL detection combined with HTTP follow-up response behavior. Make up the drawbacks that the simple TTL detection cannot identify correctly the HTTP hijacking behavior under the condition of TTL tampering. Let the False Positive Ratio reduced to lowest. 
\subsubsection{collaborative detection} Using the theory of collaborative detection can accurately locate the point of the spectral hijacking , so then, the significant and difficult problem in spectral hijacking and in detecting and locating has been solved. And the traceability success rate increased to 90
\subsubsection{share the cost} Each participant is only responsible for monitoring their own BRAS domain. In this way, the monitoring cost of the whole system is shared, so there is no high cost and scalability problems faced by detection locating system. However, the work of this paper can not replace the operators’ full network hijacking detection locating scheme. We did not take HTTPS and hijacking attack point located below the BRAS device into account the hijacking.

\section{Hijack detection deployment topology}
\begin{figure}[ht]
	\centering
	\begin{tabular}{cc}
		\includegraphics[width=2.8 in]{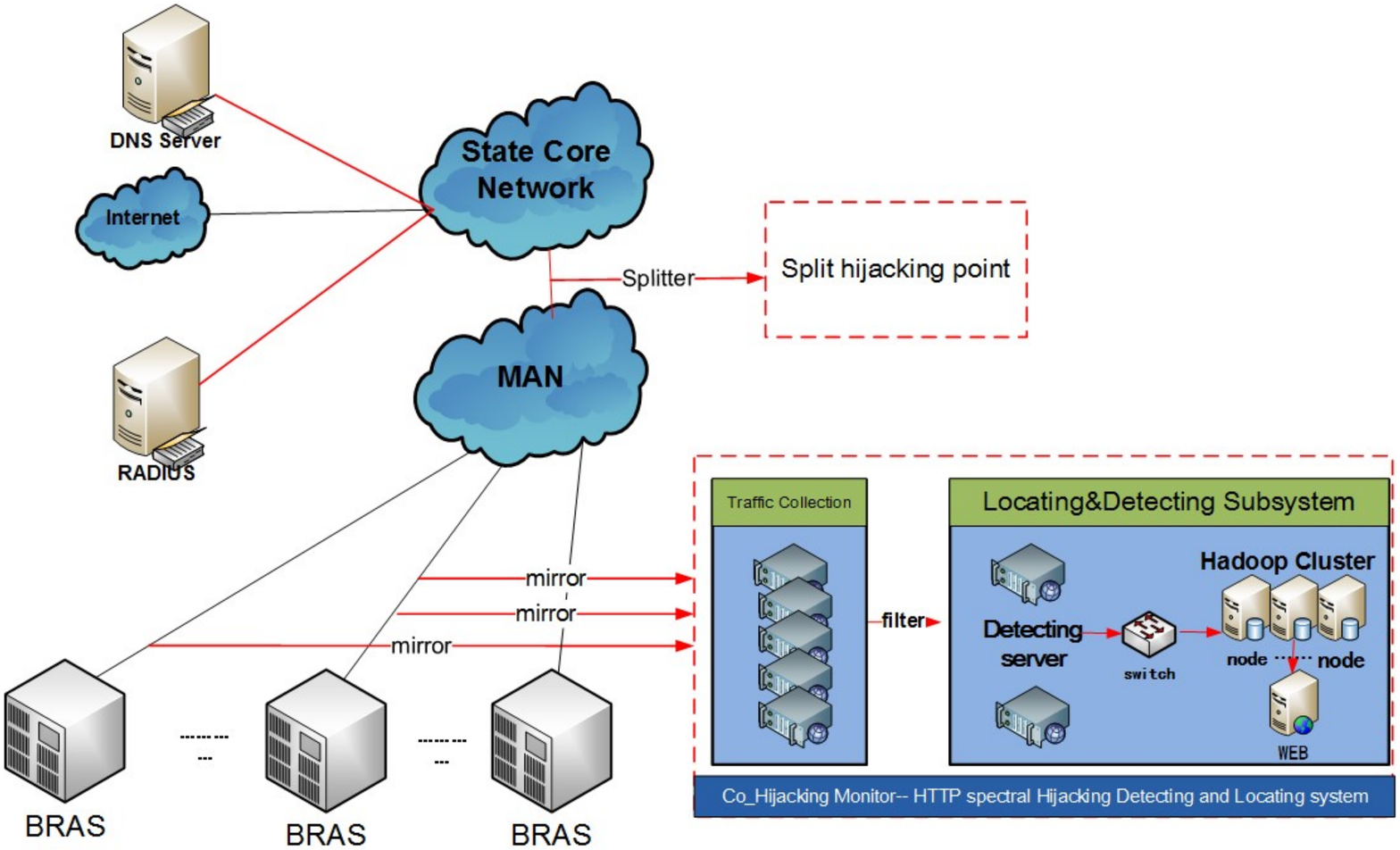}
	\end{tabular}
	\caption{HTTP spectral hijacking detection system deployment}\label{fig:Figure6detectdeployment}
\end{figure}

Several conditions are needed to meet the deployment of the spectral detection system. First, the hijacking detection point must be close enough to the user side. The purpose is to capture the interference of the camouflage response packet flow. Second, the deployment costs should be low enough. If the deployment cost is too large, it does not have the implementation of feasibility.

In this paper, the detection system is divided into two parts, mainly including the traffic collecting device and locating subsystem, as shown in Fig.~\ref{fig:Figure6detectdeployment}. 
\subsubsection{traffic collection} Traffic collecting point is deployed in the operator's access nodes adjacent to the BRAS device. 
This is mainly to take into account the BRAS device access to almost all of the Internet terminal, at the same time the device is located in the access authentication level, close enough to the user, the spectral hijacking attacks are usually in the upper layer of the BRAS, which can ensure that the traffic acquisition equipment can fully capture the packet flow of the interference. 
\subsubsection{flow mirror method} From the perspective of acquisition and construction costs, flow collection methods using the flow mirror method, in order to avoid a large number of investments and construction of spectral equipment.  BRAS devices often have the ability to mirror the flow, in particular, can be directly configured to mirror the flow of a certain type of port, such as HTTP up and down flow required for this paper. 
\subsubsection{flow filtering} Traffic collection equipment mainly work for simple flow filtering. In this paper, we mainly need to filter out the HTTP response packet, such as 301, 302, 200, etc. 
\subsubsection{detecting and locating subsystem} The subsystem mainly perfom the detecting and tracking of HTTP, including the detecting server and storage node and ethernet switch. 

The detection and tracking algorithms will be described in detail below.

\subsection{Hijacking detection algorithm}
Here we define the related notation and terms. 
\subsubsection{BN}the detection area set, which is the detection area of the BRAS device. BN is a set of BRAS devices, that is
$$ BN=\{\beta_1,\beta_2,\beta_3...\beta_n\} $$
$ \beta $ refers to the BRAS devices. $ \beta \epsilon BN, 1 \leq n \leq N $, such as $ \beta_1 $ refers to BRAS1, $ \beta_n $ refers to BRASn. 
\subsubsection{SN}the set of hijacked target sites, that is
$$ SN=\{\theta_1,\theta_2,\theta_3...\theta_k\} $$
$ \theta $ refers to the host name of the target site that was hijacked, such as news.sina.com.cn. $ \theta \epsilon SN, 1 \leq k \leq K $, $ k $ is an integer, and $ K $ is the number of target sites to be detected. 
\subsubsection{IN}the IP set of the target site server that was hijacked, that is
$$ IN=\{\gamma_1,\gamma_2,\gamma_3...\gamma_j\} $$
$ \gamma \epsilon IN, 1 \leq j \leq M $, $ j $ is an integer, $ M $ is the number of the target site server IP. Considering that the target site with the same host name may have a different number of server IP addresses. 
\subsubsection{HN}the normal route hop count of the target site, that is
$$ HN=\{h_1,h_2,h_3...h_j\} $$
$ 1 \leq j \leq M $, $ h $ refers to the normal route hops of the target site.   $ j $ is an integer, $ M $ is the number of the target hijacked server IP. 
\subsubsection{UN}the actual routing hops of the target site, that is
$$ UN=\{\mu_1,\mu_2,\mu_3...\mu_i\} $$
$ \mu \epsilon UN, 1 \leq i \leq I $, $ u $ refers to the actual route number of the target site. $ i $ is an integer and $ I $ is the number of HTTP session response packets.

\begin{figure}[ht]
	\centering
	\begin{tabular}{cc}
		\includegraphics[width=3.2 in]{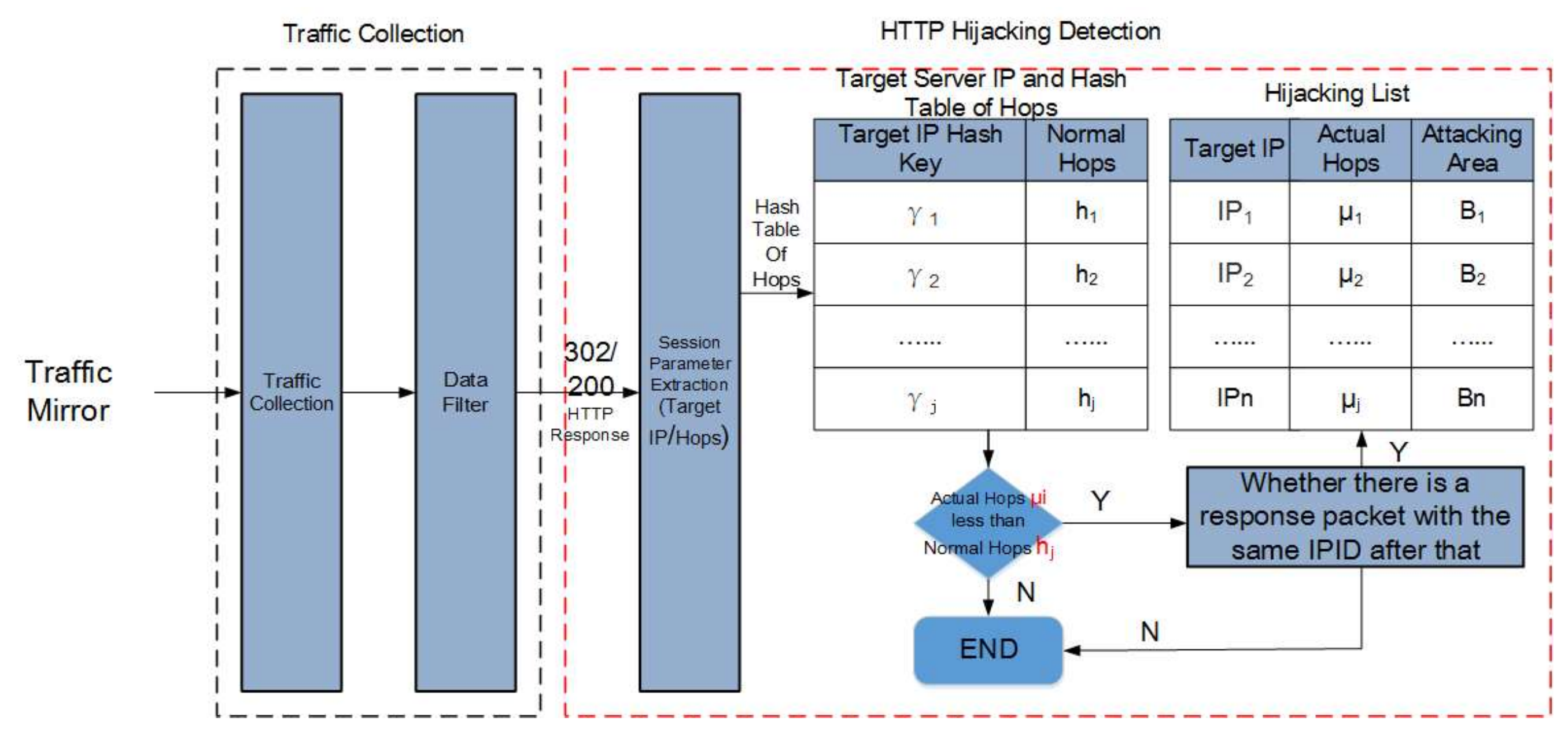}
	\end{tabular}
	\caption{Co\_HijackingMonitor detection mechanism algorithm}\label{fig:Figure7detectflow}
\end{figure}

First, the mechanism is modeled from the TTL value, or the number of routing hops, and the subsequent behavior of the normal response of the HTTP.  Study its principle, dig out the differences of HTTP spectral hijacking and normal HTTP session behavior.  Meanwhile, we established a hash table of large-scale dynamic updating routing for a set of target sites to detect TTL value of each HTTP session.  At the same time, for the part of the attacker may use tamper TTL values to avoid detection, use HTTP follow-up response behavior to further determine whether it is hijacked.  As shown in Fig.~\ref{fig:Figure7detectflow}.  Third, in order to cover as much as possible the scope of attack detection, select operator network access layer BRAS level to carry out traffic spectral detection, to set up a whole set of the whole network spectral monitoring system, detect HTTP spectral hijacking efficiently and quickly. 

The detection algorithm is as follows:\\
(1) Build the hash table of a normal number of routes in the Top N site based on TTL value. Considering the huge number of the target site, the number of routing tables will be very great. however, it is easy to search and find to store those tables by hash table. The target site server IP address can be the index of hash table, as shown in Table 1. 
\begin{table}[ht!]
	\centering
	\caption{THE TARGET SITE ROUTING HOPS TABLE}\label{table:table_1}
	\begin{tabular}{|p{0.5cm}|p{1.4cm}|p{1.4cm}|p{1.4cm}|p{1.4cm}|}
		\hline
		No & SN & IN & BN & HN \\
		\hline
		1  & $ \theta_1 $ & $ \gamma_1 $ & $ \beta_1 $ & $ h_1 $\\
		\hline 
		2  & $ \theta_2 $ & $ \gamma_2 $ & $ \beta_2 $ & $ h_2 $\\
		\hline 
		3  & $ \theta_3 $ & $ \gamma_3 $ & $ \beta_3 $ & $ h_3 $\\
		\hline 
		...  & ... & ... & ... & ...\\ 
		\hline
		n  & $ \theta_n $ & $ \gamma_j $ & $ \beta_n $ & $ h_j $\\ 
		\hline
	\end{tabular}	
\end{table}

(2) Traffic collection, filtering, extraction of session parameters. 
Co\_HijackingMonitor is mainly aimed at the HTTP response, especially the HTTP 302 redirect response and the 200 OK response to the session parameters extraction, the extracted information mainly includes the target server IP address, TTL value and so on. 

(3) comparing route hop count. 
It is mainly to compare the extracted session parameters with the target site routing hops hash table.  If the number of is less than the normal, the session will be marked as suspicious. 

(4) Determine whether there is the same serial number of the response packet in this session or not. If there is, it is determined that the session has been hijacked, otherwise, the suspicious mark of the session is cleared. 

(5) record all spectral hijacking information, as shown in Table 2.  
The list contains the target site server IP address, attack area, IP address of victim and other information. 
\begin{table}[ht!]
	\centering
	\caption{hijacking detail record list TABLE}\label{table:table_1}
	\begin{tabular}{|p{0.5cm}|p{1cm}|p{1cm}|p{1cm}|p{2.6cm}|}
		\hline
		No & SN & IN & BN & IP Address of Victim \\
		\hline
		1  & $ \theta_1 $ & $ \gamma_1 $ & $ \beta_1 $ & $ IP_1 $\\
		\hline 
		2  & $ \theta_2 $ & $ \gamma_2 $ & $ \beta_2 $ & $ IP_2 $\\
		\hline 
		3  & $ \theta_3 $ & $ \gamma_3 $ & $ \beta_3 $ & $ IP_3 $\\
		\hline 
		...  & ... & ... & ... & ...\\ 
		\hline
		n  & $ \theta_n $ & $ \gamma_j $ & $ \beta_n $ & $ IP_j $\\ 
		\hline
	\end{tabular}	
\end{table}

\subsection{Detecting and locating algorithm}
\begin{figure}[ht]
	\centering
	\begin{tabular}{cc}
		\includegraphics[width=3.2 in]{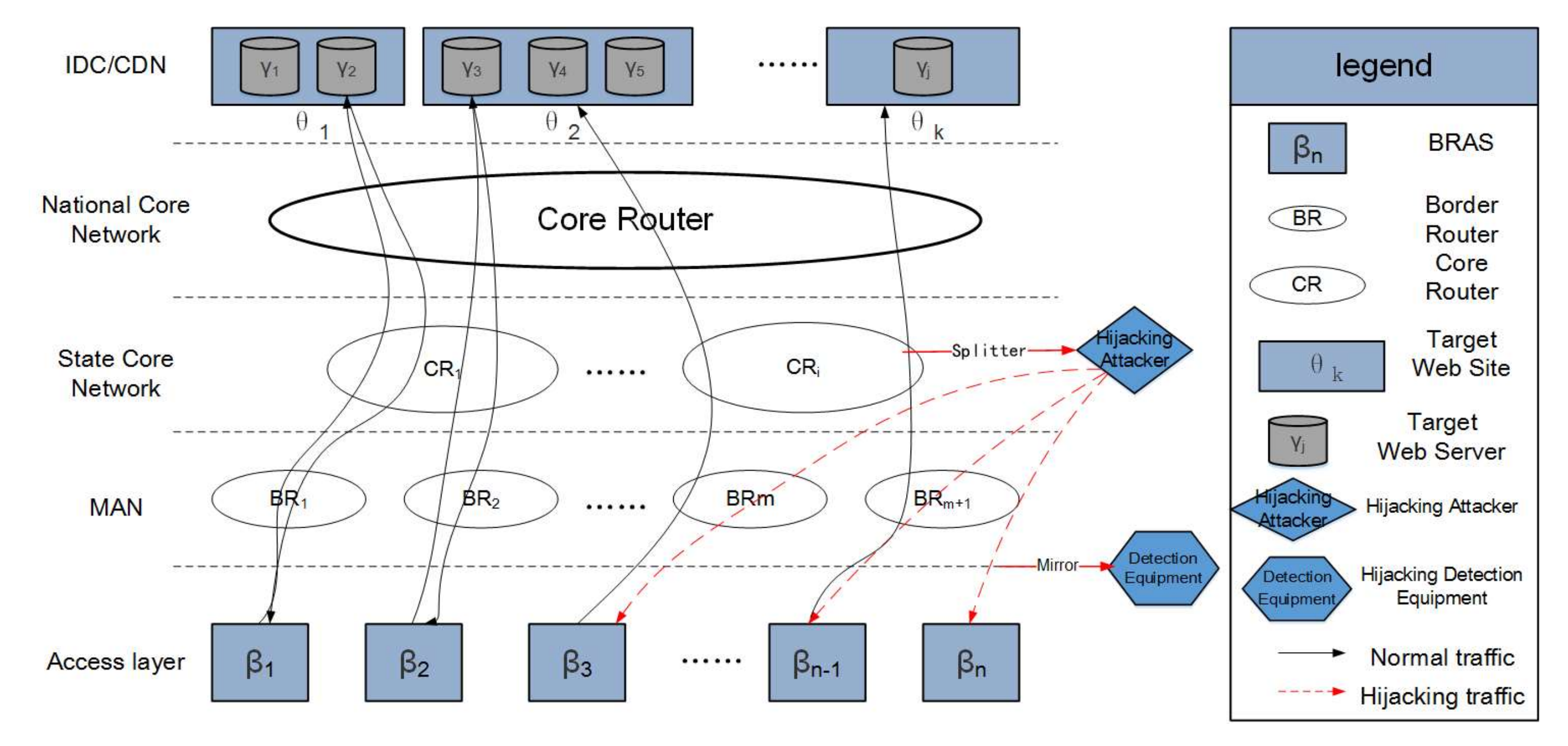}
	\end{tabular}
	\caption{Co\_HijackingMonitor attack detecting and locating algorithm}\label{fig:Figure8detectalgorithm}
\end{figure}

It is also very important to locate or trace the hijacking attack event. 
We should locate the attack point at which level of the network, such as access, convergence, core network and so on. The attack target is often not targeted by the attackers, that is, the spectral hijacking attacker will not be specific to the broadband Internet users. Therefore, the attack behavior will be irregularly scattered in different BRAS area BN, which makes the location of the source of the attack is very difficult.  But the base of the attacker's attack is the bypass spectral, so the attack point is often collected in all the HTTP traffic of a certain level network.  As shown in Fig.~\ref{fig:Figure8detectalgorithm}. 

The spectral hijacking tracing algorithm is as follows:\\
(1) create a routing topology table based on network level, as shown in Table 3. \\
(2) Find out the statistics of attack area BN’s distribution principle.  That is
$$ BN=\{\beta_1,\beta_2,\beta_3...\beta_n\} $$
(3) Step up the network level one by one, estimate the convergence point of upper network level in the area of $ BN $. \\
(4) Repeat step (2), until convergence points to a single routing device or a routing autonomous domain. So the hijacking attack occurred in the router. 

\begin{table}[ht!]
	\centering
	\caption{ROUTING TOPOLOGY OF NETWORK LEVEL}\label{table:table_1}
	\begin{tabular}{|p{0.5cm}|p{1.5cm}|p{0.5cm}|p{1.5cm}|p{1.5cm}|p{0.5cm}|}
		\hline
		No & IP Address of Victim & BN & Border Router & Core Router & IN\\
		\hline
		1  & $ IP_1 $  & $ \beta_1 $ & $ BR_1 $ & $ CR_1 $ & $ \gamma_1 $ \\
		\hline 
		2  & $ IP_2 $  & $ \beta_2 $ & $ BR_2 $ & $ CR_2 $ & $ \gamma_2 $ \\
		\hline 
		3  & $ IP_3 $  & $ \beta_3 $ & $ BR_3 $ & $ CR_3 $ & $ \gamma_3 $ \\
		\hline 
		...  & ... & ... & ... & ... & ...\\ 
		\hline
		n  & $ IP_n $  & $ \beta_n $ & $ BR_n $ & $ CR_n $ & $ \gamma_n $ \\
		\hline
	\end{tabular}	
\end{table}

\begin{figure}[ht]
	\centering
	\begin{tabular}{cc}
		\includegraphics[width=3.2 in]{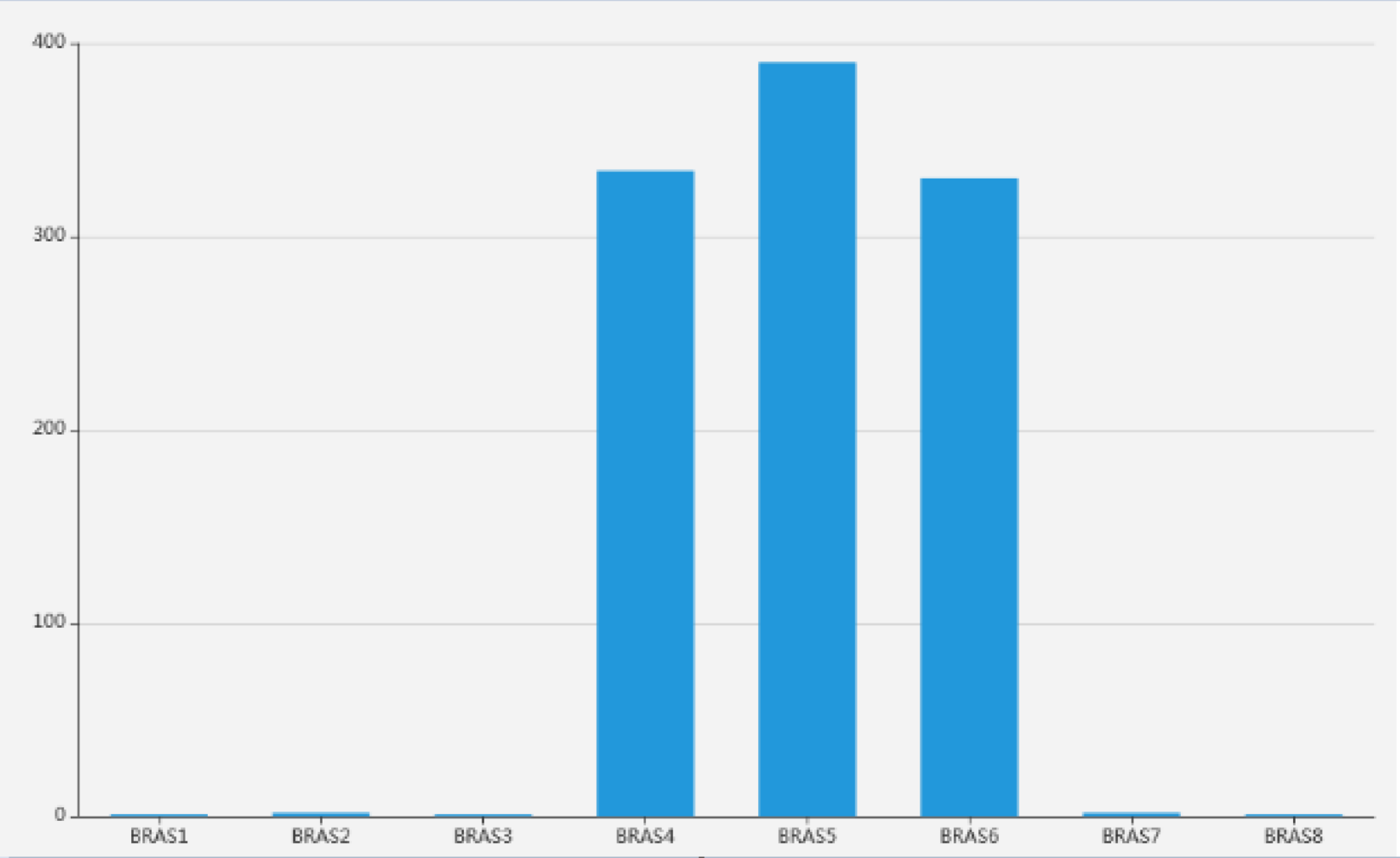}
	\end{tabular}
	\caption{the distribution of the BRAS area of the hijacking attack}\label{fig:Figure9distribution}
\end{figure}

\begin{figure}[ht]
	\centering
	\begin{tabular}{cc}
		\includegraphics[width=2.8 in]{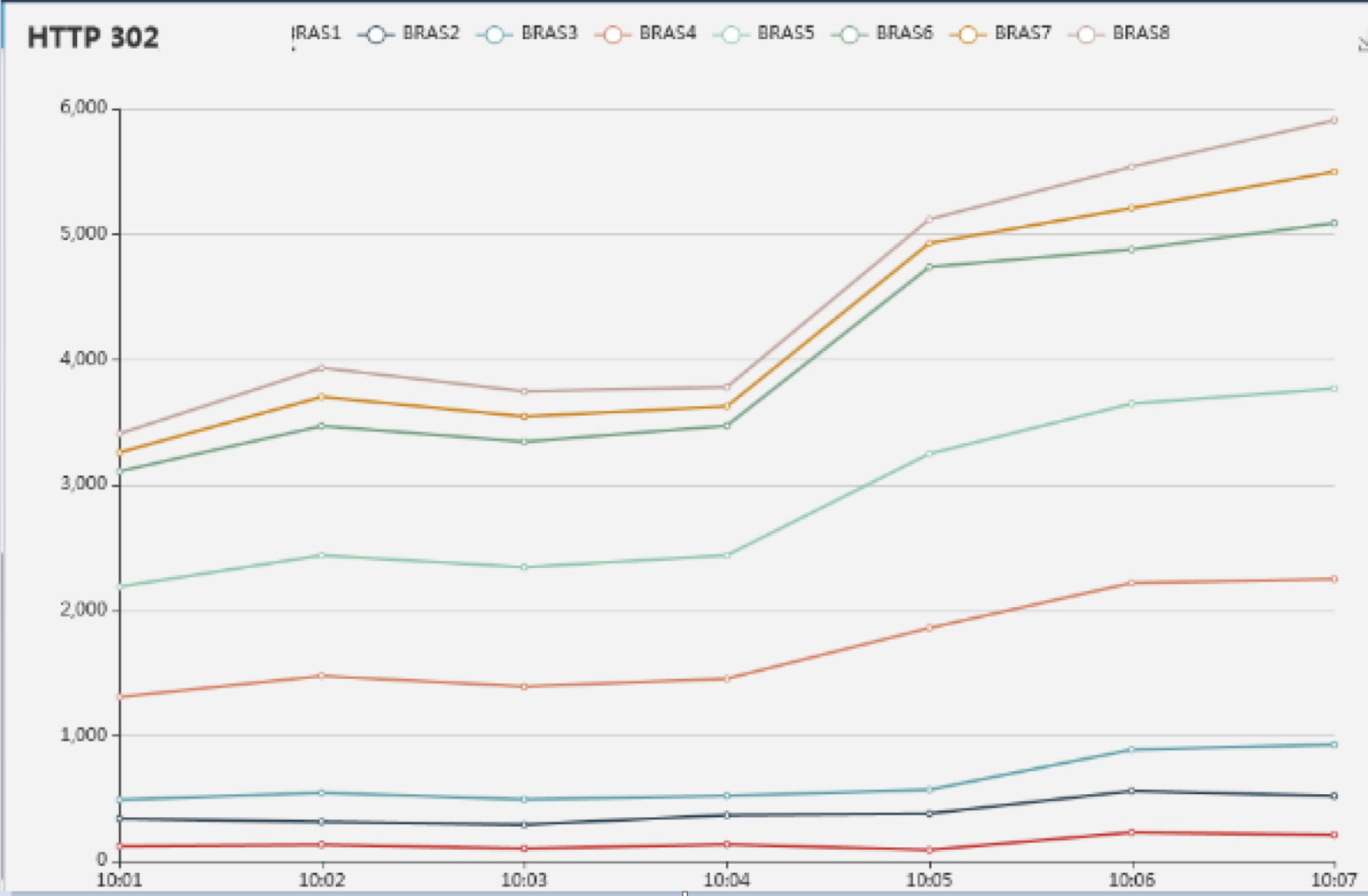}
	\end{tabular}
	\caption{Response distribution of HTTP 302 in different BRAS areas}\label{fig:Figure10hopcurve}
\end{figure}

As what can be seen from Figure.~\ref{fig:Figure9distribution}, the distribution of points of spectral hijacking attacks in some specific BRAS devices.  At the same time, according to table 3, we can follow it to find these specific BRAS access devices are concentrated on one BR.  And at the same time, by comparing the distribution of the HTTP 302 redirection response on different BRAS, as shown in Figure.~\ref{fig:Figure10hopcurve}, it is also very obvious that these attacks occur on the specific BRAS devices received on the HTTP 302 redirection response accounted for an unusually high.  It further proves and confirms the occurrence of these attacks and the access point of the hijacking attack. 


\section{EXPERIMENT AND EVALUATION}
Usually, people evaluate the detection capability of the system by two key indicators, which are mistake rate and miss rate.  The basic indicators are defined:True Positive (TP), represents the number of samples that are correctly detected as hijacking attacks, and the False Positive (FP), represents the number of samples that are mistakenly detected as hijacking attacks. True Negative (TN), represents the number of samples correctly detected for normal sessions.  False Negative (FN), represents the number of samples mistakenly detected for normal sessions.  From this, we can define a number of commonly used indicators. 
\subsection{Detection Rate}that is the True Positive Rate, sometimes referred to as the Recall Rate, represents the proportion of samples that are correctly detected as hijacking attacks in all hijacked samples. \subsection{False Alarm Rate}that is False Positive Rate, represents the proportion of samples that are mistakenly detected as hijacking attacks in all hijacked samples. 
\subsection{Missed Detection Rate}that is False Negative Rate, represents the proportion of samples that are mistakenly detected as normal sessions. 
\subsection{Classification accuracy (ACC)} is the ratio of all correctly detected samples to all samples.  That is:
$$ ACC = (TP+TN)/(TP+TN+FP+FN) $$

\begin{figure}[ht]
	\centering
	\begin{tabular}{cc}
		\includegraphics[width=2.8 in]{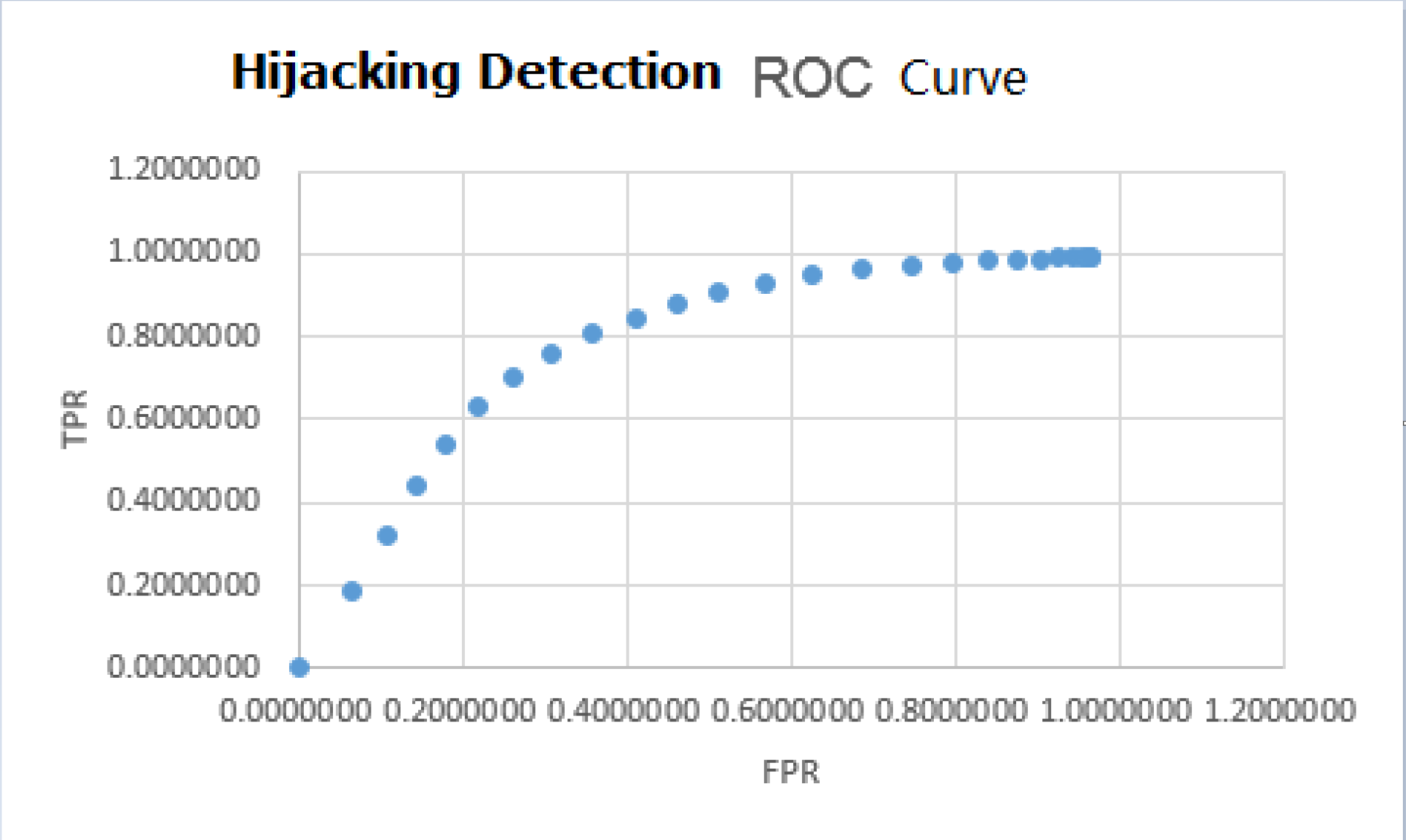}
	\end{tabular}
	\caption{ ROC curve of the accuracy of the detection of spectral hijacking}\label{fig:Figure11ROC}
\end{figure}

We set up a real HTTP man in the middle attack environment, access spectral related hijacking attacks device in the BR side of an operator's metropolitan area network.  We use multi frequency, a short period of hijacking attacks in order not to affect the user's normal Internet access, as far as possible to reduce the user's perception. 
We have organized nearly 30 measurements, each time the attack time is measured in 5 minutes, from the ROC curve of Figure.~\ref{fig:Figure11ROC}, we can see that, in this paper, the accuracy of the Co\_Hijacking Monitor detection algorithm is very high, close to $ 99\% $. 

\section{Conclusion and Future Work}\label{sec:conclusion}
Attack and defense in network have been the driving force of the entire the network security and network circles.  Only by constantly discovering the weakness of the network, we can  establish and improve the security and robustness of the network. Aiming at the spectral hijacking which has been a social realistic problem, we put forward the detection and localization of spectral hijacking.  The benefits of ISP have played a positive role in improving operator pipeline safety and protecting users.
However, with the popularity of HTTPS, the difficulty of network hijacking detection will increase, the detection and prevention of this problem, it will be the focus of our next step.

\renewcommand\refname{Reference}
\bibliographystyle{IEEEtran}
\bibliography{IEEEfull,Reference}

\begin{thebibliography}{10}
\providecommand{\url}[1]{#1}
\csname url@samestyle\endcsname
\providecommand{\newblock}{\relax}
\providecommand{\bibinfo}[2]{#2}
\providecommand{\BIBentrySTDinterwordspacing}{\spaceskip=0pt\relax}
\providecommand{\BIBentryALTinterwordstretchfactor}{4}
\providecommand{\BIBentryALTinterwordspacing}{\spaceskip=\fontdimen2\font plus
\BIBentryALTinterwordstretchfactor\fontdimen3\font minus
  \fontdimen4\font\relax}
\providecommand{\BIBforeignlanguage}[2]{{%
\expandafter\ifx\csname l@#1\endcsname\relax
\typeout{** WARNING: IEEEtran.bst: No hyphenation pattern has been}%
\typeout{** loaded for the language `#1'. Using the pattern for}%
\typeout{** the default language instead.}%
\else
\language=\csname l@#1\endcsname
\fi
#2}}
\providecommand{\BIBdecl}{\relax}
\BIBdecl

\bibitem{browserhijacking}
N.~Mazher, I.~Ashraf, and A.~Altaf, ``Which web browser work best for detecting
  phishing,'' in \emph{2013 5th International Conference on Information and
  Communication Technologies}, Dec 2013, pp. 1--5.

\bibitem{httpproxy}
T.~D. Laksono, Y.~Rosmansyah, B.~Dabarsyah, and J.~U. Choi, ``Javascript-based
  device fingerprinting mitigation using personal http proxy,'' in \emph{2015
  International Conference on Information Technology Systems and Innovation
  (ICITSI)}, Nov 2015, pp. 1--6.

\bibitem{kernelhijacking}
X.~Li, Y.~Zhang, and Y.~Tang, ``Kernel malware core implementation: A survey,''
  in \emph{2015 International Conference on Cyber-Enabled Distributed Computing
  and Knowledge Discovery}, Sept 2015, pp. 9--15.

\bibitem{bootkit}
H.~Gao, Q.~Li, Y.~Zhu, W.~Wang, and L.~Zhou, ``Research on the working
  mechanism of bootkit,'' in \emph{2012 8th International Conference on
  Information Science and Digital Content Technology (ICIDT2012)}, vol.~3, June
  2012, pp. 476--479.

\bibitem{cachepollution}
R.~C. M and G.~P. Sajeev, ``Intelligent pollution controlling mechanism for
  peer to peer caches,'' in \emph{2015 Seventh International Conference on
  Computational Intelligence, Modelling and Simulation (CIMSim)}, July 2015,
  pp. 141--146.

\bibitem{kong2014}
Z.~Kong, ``Dns spoofing principle and its defense scheme,'' \emph{Computer
  Engineering}, vol.~36, no.~3, pp. 107--110, 2013.

\bibitem{chongqing}
\text{Zhao ZiDong}, ``Chongqing telecom employees who hijacked the user traffic
  for personal gain millions were arrested,'' \emph{Available from
  \url{http://www.cq.xinhuanet.com/2015-05/25/c_1115397182.htm}}.

\bibitem{yanboru2016}
B.~Yan, ``Detection and prevention of dns spoofing attack,'' \emph{Computer
  Engineering}, vol.~32, no.~21, pp. 130--132, 2016.

\bibitem{wangpeng2013}
P.~Wang, ``Research on the hijacking of http session in switched network and
  its policy,'' \emph{Computer Engineering}, vol.~33, no.~5, pp. 135--137,
  2013.

\bibitem{yangbo2015}
B.~Yang, ``A locating method for splitting hijacking interference,''
  \emph{Information Security and Technology}, vol.~6, no.~12, pp. 70--74, 2015.

\bibitem{Hajikarami2014}
F.~Hajikarami, M.~Berenjkoub, and M.~H. Manshaei, ``A modular two-layer system
  for accurate and fast traffic classification,'' in \emph{2014 11th
  International ISC Conference on Information Security and Cryptology}, Sept
  2014, pp. 149--154.

\bibitem{Meda2014}
C.~Meda, F.~Bisio, P.~Gastaldo, and R.~Zunino, ``A machine learning approach
  for twitter spammers detection,'' in \emph{2014 International Carnahan
  Conference on Security Technology (ICCST)}, Oct 2014, pp. 1--6.

\bibitem{Mauro2014}
M.~D. Mauro and M.~Longo, ``Skype traffic detection: A decision theory based
  tool,'' in \emph{2014 International Carnahan Conference on Security
  Technology (ICCST)}, Oct 2014, pp. 1--6.

\end{thebibliography}

\end{document}